\documentclass[a4paper]{mem}
\usepackage{natbib}
\usepackage{graphicx}
\usepackage[a4paper]{hyperref}
\idline{???}{1}
\begin{document}
   \title{Simulating Galaxies and Galaxy Clusters
}

   \author{J.\ Sommer-Larsen$^{1,2}$}


   \institute{$^1$Theoretical Astrophysics Center, Juliane Maries Vej 30,
2100 Copenhagen \O \\
$^2$Nordita, Blegdamsvej 17, 2100 Copenhagen \O\\  
 \email{jslarsen@tac.dk} 
             }

   \abstract{The X-ray luminosities of the hot halo gas around simulated, 
Milky Way
like disk galaxies have been determined, as a function of redshift. The X-ray
luminosity increases significantly with redshift, in some cases as much
as a factor 30 going from $z$=0 to 2. Consequently, the optimal detection
redshift can be $\ga$0.5. 

Results of fully cosmological simulations of
galaxy groups and clusters, incorporating star formation, chemical evolution
with non-instantaneous recycling, metal-dependent radiative cooling,
galactic super-winds and thermal conduction are presented. X-ray luminosities
at $z$=0 are somewhat high and central entropies somewhat low compared to
observations. This is likely a combined effect of chemical evolution
and metal-dependent, radiative cooling. Central ICM abundance profiles are
somewhat steep, and the observed level of ICM enrichment can only be 
reproduced with IMFs more top-heavy than the Salpeter IMF. In agreement with
observations it is found that the iron in the ICM is in place already at
$z\sim$1. The [Si/Fe] of the ICM decreases with time, and
increases slightly with radius at a given time. The cluster galaxies match
the observed ``red sequence'' very well, and the metallicity of
cluster galaxies increases with galaxy mass, as observed.

   \keywords{Numerical simulations - Galaxies - Clusters
	     of galaxies}
   }
   \authorrunning{J.\ Sommer-Larsen}
   \titlerunning{Simulating galaxies and galaxy clusters}
   \maketitle
%


\section{X-ray emission from the haloes of disk galaxies}
Standard models of disk galaxy formation require spiral galaxies to be 
surrounded by large reservoirs of hot gas, which should be emitting at X-ray 
wavelengths, and from which gas should still be accreting on to the disk at
present (e.g., White \& Frenk 1991).
The detailed properties of these haloes and their role in the 
formation and evolution of galactic disks remain largely unknown from an
observational viewpoint, as the haloes have so far escaped direct detection.
Possible exceptions are NGC891 (Bregman \& Houck 1997; Strickland et al. 
2004), NGC 2841 (Benson et al. 2000), 
and the Milky Way itself (e.g., Sidher et al. 1996; Pietz et al 1998), 
if neglecting cases where (i) the 
galactic disks show evidence of being disturbed by tidal interactions, or (ii)
the X-ray emission at off-disk distances of a few kpc can be attributed to 
processes originating in the disk such as feedback from star-bursts.
\begin{figure}
\centering
\resizebox{0.5 \textwidth}{!}{{\includegraphics{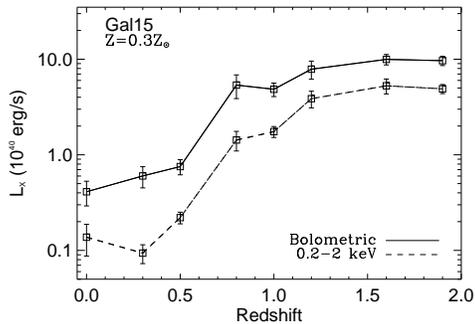}}}
\caption{Redshift evolution of the rest-frame bolometric and 0.2--2 keV X-ray
luminosities of a simulated, Milky Way like disk galaxy.}
\label{fig1}
\end{figure}

\vspace{-4mm}
Toft et al. (2002) extracted
disk galaxies from fully cosmological hydrodynamical simulations 
and determined their present-day halo X-ray luminosities, finding 
consistency with observational upper limits on halo emission from nearby 
spirals. 
Moreover, based on the predicted rates of mass accretion onto the 
disks, they suggested that spiral galaxy haloes
were
possibly one order of magnitude brighter in soft X-rays at $z\sim1$ than at
present (the gas mass cool-out rate is expected to be proportional to the halo
X-ray luminosity times the emissivity weighted inverse temperature of the
hot halo gas --- see Rasmussen et al. 2004 for details).
Analytical models of halo emission overpredict 
emission at $z=0$ by at least an order of magnitude (see Benson et al. 2000 and
Toft et al. 2002) and thus cannot be assumed to provide a reliable 
description of the evolution in X-ray emission with redshift.
Rasmussen et al. 2004 extended the work of Toft et al. by studying the 
higher-redshift properties of two Milky Way like disk galaxies extracted from 
the $\Lambda$CDM galaxy formation simulations of Sommer-Larsen, G\"otz \& 
Portinari 2003. Figure 1 shows the redshift evolution of the (rest-frame)
bolometric and 0.2--2 keV (X-ray) luminosity of the halo of a Milky Way like 
disk galaxy. The simulation was
run with a radiative cooling function corresponding to 1/3 solar abundance.
The increase in luminosity is about a factor of 30 going from $z$=0 to 2.
Given this, it is obviously of interest to determine the optimal redshift for 
X-ray halo detection, taking redshifting of the photons, cosmological
$(1+z)^4$ surface brightness dimming, and Galactic HI absorption into account.
Rasmussen et al. find that it varies from galaxy to galaxy, but in some
cases the optimal detection redshift can be $\ga$0.5. Figure 2 shows for one 
such case (the galaxy from Fig.~1) the predicted, edge-on 0.3-2 keV surface 
brightness profiles at various redshifts. An observer at the current epoch
would measure the largest surface brightness of this galaxy at 
$|\rm{z}|$$\la$10 kpc, when it was at $z\sim$0.8.    

\begin{figure}
\centering
\resizebox{0.5 \textwidth}{!}{{\includegraphics{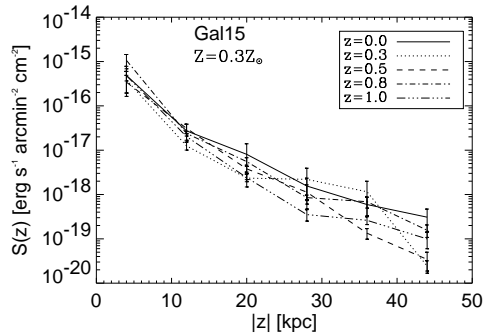}}}
\caption{0.3--2 keV surface brightness of the galaxy from Fig.~1, 
averaged over 40 kpc wide slabs oriented parallel the disk, 
as a function of vertical distance $|\rm{z}|$ to the disk. The galaxy is
viewed edge-on; for each vertical bin, the values above and below the disk 
have been added.}
\label{fig2}
\end{figure}

\section{Simulating galaxy groups and clusters --- the code and the 
simulations}
Only fairly recently has it been possible to carry out fully cosmological 
gas-dynamical/N-body simulations of the formation and evolution of galaxy
clusters at a level of numerical resolution and physical sophistication that 
the cool-out of gas, star-formation and chemical evolution in, and outflows 
from, 
individual cluster galaxies can be modelled to, at least some, degree of 
realism (e.g., Valdarnini 2003; Tornatore et al. 2004).

We have recently engaged in undertaking such simulations of galaxy groups and 
clusters. Scientific objectives include reproducing the observed 
luminosity--temperature relation for groups and clusters and its 
evolution with redshift; surface brightness and
entropy profiles; and the high iron abundance of $\sim$1/3 solar in the
intracluster medium (ICM). The metals have to be first produced in the
galaxies and subsequently expelled and mixed into the ICM;
since the mass of stars in galaxies is only about 10-20\% of the mass of
hot ICM gas, this requires high efficiency in metal production and expulsion.
Moreover one should reproduce observed ICM abundance profiles and 
$\alpha$--to--iron ratios, ``cold fractions'', and  
group and cluster galaxy luminosity functions and ``red sequences''.

The code used for the simulations is a significantly improved version of
the TreeSPH code used previously for galaxy formation simulations 
(Sommer-Larsen, G\"otz and Portinari 2003): Firstly, we have adopted the ``conservative'' entropy eq. solving scheme suggested
by Springel \& Hernquist (2002). Secondly, non-instantaneous chemical 
evolution tracing
9 elements (He, C, N, O, Mg, Si, S, Ca and Fe) has been incorporated
in the code following Lia, Portinari \& Carraro (2002). 
Atomic radiative cooling depending
both on the metal abundance of the gas and the meta-galactic UV field
is invoked together with simplified radiative transfer. Thirdly, continuing,
star-burst and/or AGN driven galactic super-winds are incorporated in the 
simulations.
This is required to get the abundance of the ICM to the observed level
of about 1/3 solar in iron. The strength of the super-winds is modelled
through a free parameter $f_{\rm{wind}}$ which determines how large a fraction
of the stars born partake in super-wind driving star formation. We find that
in order to get an iron abundance in the ICM comparable to observations,
$f_{\rm{wind}}\ga0.5$ and at the same time a fairly top-heavy Initial Mass
Function (IMF) has to be used --- see section 3. Finally, 
thermal conduction was implemented in the code following Cleary \& Monaghan
(1999).

\begin{figure}
\centering
\resizebox{0.5 \textwidth}{!}{\includegraphics{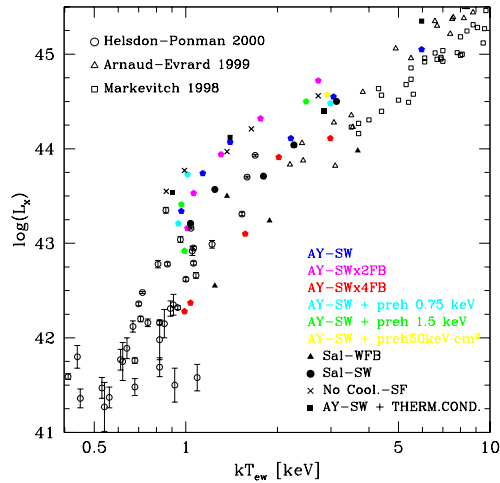}}
\caption{$L_X-T_X$ relation at $z$=0. Data points are marked by open symbols,
simulations by filled symbols and crosses. Details about the simulations are
given in the text.}
\label{fig3}
\end{figure}

Six groups and clusters ($T$=1-6 keV) were selected from a cosmological, 
$L$=150 $h^{-1}$Mpc, 
``standard'' $\Lambda$CDM N-body simulation, and re-simulated with baryonic
physics included, using the code described above, for a suite of 
physical parameters. 
The wind strength parameter $f_{\rm{wind}}$ was set to 0.7 and
Salpeter (Sal) and Arimoto-Yoshii (AY) IMFs were adopted. Some 
simulations
were run with galactic super-wind feedback two times (SWx2FB) and four
times (SWx4FB) as energetic as is available from supernovae; the remaining
amount of energy is assumed to come from AGN activity. Others were run
with an AY IMF, strong feedback (SW) and additional preheating at $z$=3
(AY-SW + preh.),
as discussed by Tornatore et al. (2003). One series of
simulations was run with a Salpeter IMF and only early ($z$$\ga$4), strong 
feedback, as in the galaxy formation simulations of  Sommer-Larsen, 
G\"otz and Portinari (2003) (as this results in overall fairly weak
feedback we denote this: Sal-WFB). Another was run with thermal 
conduction
included (AY-SW + Therm.Cond.), assuming a conductivity of 1/3 of the 
Spitzer value (e.g., Jubelgas et al. 2004). Finally, for comparison with
previous work, a series of simulations was run without radiative cooling, 
star formation and thermal conduction. 

\begin{figure}
\centering
\resizebox{0.5 \textwidth}{!}{\includegraphics{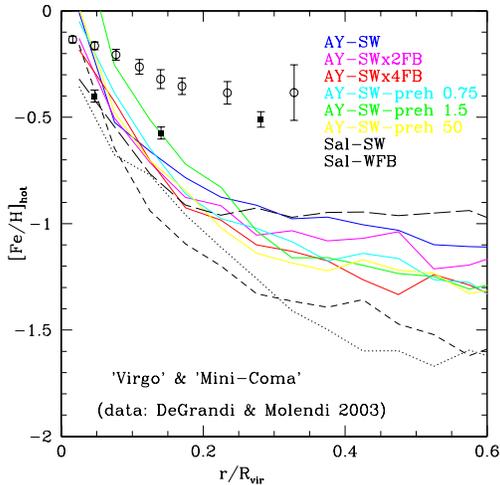}}
\caption{ICM iron abundance profiles for various simulations of the ``Virgo''
cluster, and one (AY-SW) of the ``Mini-Coma'' cluster (long-dashed line).
Also shown are data points from De Grandi et al. (2003) for cool-core
clusters (open circles with errorbars) and non cool-core clusters (filled
circles with errorbars). The appropriate comparison of the models is to the
cool-core clusters.}
\label{fig4}
\end{figure}

\begin{figure}
\centering
\resizebox{0.5 \textwidth}{!}{\includegraphics{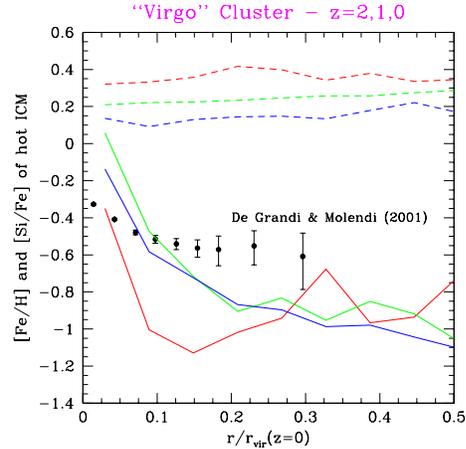}}
\caption{ICM iron abundance profiles for the AY-SW simulation of the ``Virgo''
cluster at $z$=0, 1 and 2 (solid lines). Also shown are [Si/Fe] profiles at
$z$=0, 1 and 2 (dashed lines). To the right $z$=2,1,0 corresponds to going
top down for the two sets of curves ($z$=2: red; $z$=1: green; $z$=0: blue).}
\label{fig5}
\end{figure}

\section{Simulating galaxy groups and clusters --- results}
In Figure 3 is shown the $z$=0 $L_X-T$ relation for the 6 groups and
clusters ($T$=1-6 keV), re-simulated with various physical parameters, as
explained in the previous section. 
Also shown are observational data on the $z$=0 $L_X-T$ relation from 
various
sources. The predicted X-ray luminosities are mostly somewhat high compared 
to observations, with the exception of models run with an Arimoto-Yoshii IMF
and very strong feedback (AY-SWx4FB),
and models run with a Salpeter IMF and only little feedback (Sal-WFB),
hence resulting in a low level of metal enrichment. 
The preliminary conclusion from the results of these simulations is that the 
success
of previous generations of models in reproducing the $L_X-T$ relation,
the surface brightness profiles, the entropy ``floor'' at $S=T/n_e^{2/3}\sim$
100-200 keV~cm$^2$ etc. (e.g., Tornatore et al. 2003) likely was due to the 
adoption of a primordial radiative cooling function (i.e. the contribution 
to radiative cooling from metals was neglected). The analysis of this is 
ongoing, however. For most of our models we find central entropy values of 
$\sim$ 10-20 keV~cm$^2$; this may be in agreement with ``observed''
values in cool-core clusters (M.~Voit, this conference).

\begin{figure}
\centering
\resizebox{0.5 \textwidth}{!}{{\includegraphics{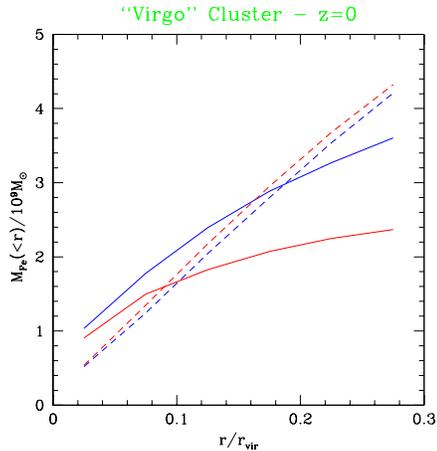}}}
\caption{Cumulative iron mass of the ICM at $z$=0 for the Sal-SW and
AY-SW simulations. Actual profiles are shown by solid lines, profiles
inferred assuming the observed iron profiles of De Grandi \& Molendi (2001)
by dashed. Top solid line corresponds to AY-SW, bottom to Sal-SW. Top
dashed line corresponds to Sal-SW, bottom to AY-SW (AY: blue; Sal: red).}
\label{fig6}
\end{figure}

\begin{figure}
\centering
\resizebox{0.5 \textwidth}{!}{{\includegraphics{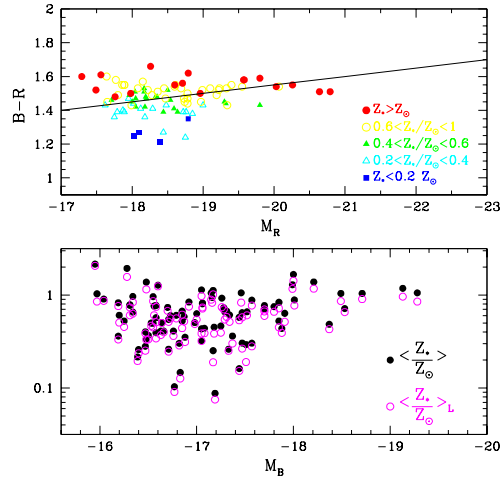}}}
\caption{{\it Top panel}: B-R vs. R Colour-Magnitude relation for the 94
`Mini-Coma' ($T$$\sim$6 keV) cluster galaxies at $z$=0, compared to the 
$z$$\simeq$0 Red Sequence of 
Gladders
et~al.\ (1998) (straight line).
{\it Bottom panel}: Metallicity--luminosity relation for the ``Mini-Coma'' 
galaxies (full symbols: mass--weigthed, open symbols: luminosity--weighted).}
\label{fig7}
\end{figure}

Figure 4 shows the profiles of iron abundance in the ICM 
for various models, together with recent observational data by De Grandi 
et al. (2003). All models, but one, were run for a $T$$\simeq$3 keV cluster,
which we shall denote ``Virgo'' in the following. One model refers to a
larger cluster with $T$$\simeq$6 keV, denoted ``Mini-Coma''. 
In general, the iron abundance profiles are too steep in the inner parts
of the clusters and too low in the outer parts. In relation to the
former problem, AGN induced, buoyant bubbles may help to transport
energy and metals from the very central part of clusters outward, 
flattening the inner abundance profiles. In relation to the latter
problem, clear progress is made going from the Salpeter IMF to the more
top-heavy Arimoto-Yoshii IMF --- see also Portinari et al. 2004. This 
indicates that an even more top-heavy IMF may be required to reproduce the
observed ICM abundance, but see also below.

Figure 5 shows, for the ``Virgo'' cluster AY-SW simulation, the profiles of
[Fe/H] and [Si/Fe] in the ICM, at redshifts
$z$=0,1 and 2. In agreement with observations the metals are essentially in 
place at $z$=1 (e.g., Tozzi et al. 2003), and it is seen that substantial 
enrichment has taken place 
already at $z$=2. This is in agreement with the findings of Sommer-Larsen et al.
(2004), that the cluster stars formed at typical redshifts
$z_f$$\ga$2.5-3. Moreover, [Si/Fe] increases slightly with radius at any
redshift, also in agreement with observations (Finoguenov et al. 2000). 
Finally, [Si/Fe] generally
decreases with time, as is expected due to the substantial, but delayed
iron enrichment of the ICM by supernovae type Ia.

Figure 6 shows the cumulative iron mass in the ICM for the Sal-SW and AY-SW
simulations of the ``Virgo'' cluster. Also shown is the cumulative mass of
iron assuming the observed ICM iron profile of De Grandi \& Molendi (2001)
(similar to the updated profiles of De Grandi et al. 2003). It is clear that
the ICM can not be sufficiently enriched for simulations adopting a Salpeter
IMF, whereas this can marginally be accomplished by using an Arimoto-Yoshii
IMF (at least out to $r$$\sim$0.2-0.3$r_{\rm{vir}}$). 

Concerning the properties of the cluster galaxies, shown in Figure 7, at $z$=0,
is the colour-magnitude (B-R versus R) relation of the 94 galaxies in the
``Mini-Coma'' cluster (excluding the cD). For comparison is also shown the 
mean locus of the observed red sequence (solid line) from Gladders et al. 
(1998). The match to observations is quite satisfactory. Furthermore is shown 
the mean heavy element abundance for the 94 galaxies versus 
absolute B-band magnitude. The non-zero slope of the colour-magnitude relation
for our cluster galaxies mainly results from the increase of the typical
stellar metallicity with galaxy luminosity, as observed
(e.g., Kodama \& Arimoto 1997).

Full details about our results are given in Romeo, Portinari \& Sommer-Larsen 
(2004), Romeo et al. (2004) and Sommer-Larsen et al. (2004).
\vspace{-0.5cm}
\section*{Acknowledgements}
\vspace{-2mm}
I thank the organizers for an excellent conference. We have benefited from
discussions with S.~Andreon, V.~Antonuccio, M.~Arnaboldi, S.~Borgani, 
R.~Bower, A.~Brandenburg, 
F.~Calura, S.~De Grandi, K.~Freeman, M.~Hoeft, D.~Kawata, S.~Leccia,
Y.~Lin, F.~Matteucci, G.~Murante, S.~Recchi, L.~Tornatore and P.~Tozzi. 
%
\vspace{-0.5cm}
\bibliographystyle{aa}

\end{document}